\documentclass[twocolumn,amssymb, nobibnotes, showpacs, superscriptaddress, aps, prd]{revtex4}
\usepackage{amsmath}
\usepackage{epsfig}

\begin{document}
\title{Collective atomic recoil motion in short-pulse multi-matter-optical wave mixing}
\author{L. Deng}
\affiliation{Physics Laboratory, National Institute of Standards \& Technology, Gaithersburg, Maryland 20899}
\affiliation{Center for Cold Atom Physics, Wuhan Institute of Physics and Mathematics, Chinese Academy of Science, Wuhan 430071, China}
\author{E.W. Hagley}
\affiliation{Physics Laboratory, National Institute of Standards \& Technology, Gaithersburg, Maryland 20899}
\date{\today}

\begin{abstract}
An analytical perturbation theory of short-pulse, matter-wave superradiant scatterings is presented.
We show that Bragg resonant enhancement is incapacitated and both positive 
and negative order scatterings contribute equally.  We further show that propagation gain is small and scattering events primarily
occur at the end of the condensate where the generated field has maximum strength,
thereby explaining the apparent ``asymmetry" in the scattered components with respect to the condensate center.  In addition, the generated field travels near the speed of light in a vacuum, resulting in significant spontaneous emission when the one-photon detuning is not sufficiently large. Finally, we show that when the excitation rate increases, the generated-field front-edge-steepening and peak forward-shifting effects are due to depletion of the ground state matter wave.
   
\end{abstract}

\maketitle

Matter-wave superradiance is coherent and collective atomic recoil motion that was
first observed in an ensemble of ultra-cold, Bose-Einstein condensed $^{87}$Rb atoms under 
a single, long-pulsed laser excitation \cite{inouye1}. Later, Schneble {\em et al.} demonstrated 
the short pulsed, {\it bi-directional} superradiant effect \cite{schneble1}.  Many 
experimental and theoretical studies since these initial observations have provided substantial insight into this 
light-matter-wave interaction process \cite{inouye2,inouye3,moore,li,piovella,han,ketterle,bonifacio,
zobay,hilliard,uys,benedek}.  However, to date there has been no analytical theory predicting wave propagation dynamics in the short-pulse, 
matter wave superradiant scattering effect.    

In this Letter, we present an analytical small-signal propagation theory on short-pulse
matter-wave superradiant scattering.  
Five important outstanding questions are examined and answered: (1) why are $n<0$ scattering orders
not detectable under long-pump pulses yet they are readily observable when the
pump pulse is significantly shorter; (2) why is there an apparent ``asymmetry" in the atom scattering pattern with a short-pump pulse; (3) what is the cause of the significant atomic cloud and halos in short-pulse, matter-wave superradiant scattering; (4) what are the field propagation and atom scattering characteristics; and (5) what is the cause of the optical field front-edge-steepening and 
pulse-peak-forward-shifting effects when the excitation rate increases.
   
Specifically, we develop a first-order, short-pulse perturbation theory to explain optical-field generation and propagation dynamics in the context of matter-wave superradiant scattering.  Using a semi-classical approach we show that the optical field responsible for matter-wave scattering is generated by a spontaneous scattering process at the far-end of the condensate, and it reaches maximum strength after having traversed the entire length of the condensate by counter-propagating with respect to the pump field.  This physical picture clearly shows that most superradiant scattering occurs at the end of the condensate and therefore explains the apparent ``asymmetry" in the bi-directional scattering pattern in the short pulse limit.  We further show that in this limit the generated field travels with group
velocity $V_g\approx c$.  Because of the small size of the condensate this leads to a fast photon relaxation rate \cite{bonifacio,ketterle,
zobay,hilliard,dicke} that is much larger than the one-photon detuning from the upper electronic state used in all short-pulse experiments reported to date, resulting in significant population being transfered to the upper electronic state.  The consequence is severe spontaneous emission which gives rise to the atom cloud and halos observed.  Consequently, the resulting atom scattering pattern in the short-pulse limit has lost the meaning of ``coherent matter-wave scattering" because a large number of atoms are in various momentum states with random phases.
To the best of our knowledge no analytical theory to date is capable of explaining these important findings.

\begin{figure}
\centering
\includegraphics[angle=90,width=3in]{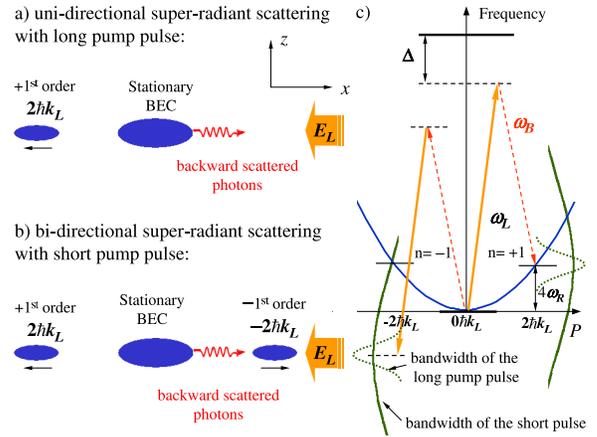} 
\caption{Schematic drawings of longitudinally-excited matter-wave superradiant scattering from a Bose condensate under long (a), and short (b) excitation pulses.  (c) Schematic drawing of the energy-momentum dispersion for a long-pump pulse (narrow band, dotted profile) and a short-pump pulse (broad band, solid profile).  Only orders $n=-1,0,+1$ are shown.}
\end{figure}

To investigate optical field generation and propagation dynamics and superradiant matter wave scattering we assume 
that a small group of photons \cite{note0a} is either generated initially by spontaneous Rayleigh scattering or by injection-seeding at the end
$x=0$. We further assume that this group of photons travels along the long axis of the condensate \cite{note0b} in the $+\hat{\bf x}$ direction, and a pump field with amplitude of $E_L$ \cite{note1} traverses the condensate in the $-\hat{\bf x}$ direction (see Figure 1).  
We consider a first-order perturbation treatment of this weak initial field.  In this limit only first-order matter-wave superradiant
scatterings are important. Thus, $|n|>1$ terms are neglected and the ground state condensate remains undepleted, i.e., $|\psi_{0}|^2\approx$ const.  If the one-photon detuning is {\it adequately} selected \cite{note0}
so that adiabatic elimination of the upper excited atomic state is applicable, then the coupled Schr$\ddot{\text{o}}$dinger equation for the $n=\pm 1$ order superradiantly scattered mean-field macroscopic wave function components $\psi_{\pm 1}$ in the interaction representation, with Doppler effects included, can be expressed as: 

\begin{eqnarray}
\frac{\partial\psi_{\pm 1}}{\partial t}&=&-(4R+\gamma_B)\psi_{\pm 1}
-ig_{0}\delta\frac{E_{B}^{(\mp)}}{E_{L}^{(\mp)}}\psi_{0}\,e^{i(4\omega_{R}\mp\Delta_{L})t}, \quad
\end{eqnarray}
where $\omega_{R}=\hbar k_L^2/(2M)$ is the one-photon recoil frequency, $g_{0}=|\Omega_{L}|^2/|\Delta|^2$, $R=g_{0}\Gamma_{0}/4$ is the single photon Rayleigh scattering rate with $\Gamma_0$ being the natural linewidth of the upper electronic state.
In addition, $\gamma_B$ is the Bragg resonance linewidth of a two-photon transition between two motion states, $\Omega_{L}=d_{12}E_{L}/\hbar$
is the pump field Rabi frequency with $d_{12}$ being the dipole transition matrix element between the ground and upper electronic states.  $E_{L}^{(+)}$ ($E_{B}^{(+)}$) is the positive frequency part of the pump (superradiantly-generated) field amplitude \cite{note1} and $\Delta_{L}=\omega_{L}-\omega_{B}$ is the frequency difference between the pump and the backward-propagating fields.

To understand the disappearance and re-emergence of the $n<0$ scattering orders in the long- and short-pump-pulse limits, we analyze the driving term on the right side of Eq. (1).  It is 
clear that for a long-pump pulse under the Bragg resonance enhancement condition (i.e., $4\omega_R=\Delta_L$\cite{note1}) the phase factor on the right side of Eq. (1) for the $n=-1$ component, i.e., $e^{i(4\omega_R+\Delta_L)\tau}$, oscillates much faster than the phase factor of $n=+1$ component which
is $e^{i(4\omega_R-\Delta_L)\tau}\approx 1$.  Consequently, the probability amplitude for the $n=-1$ 
order is small whereas the amplitude for the $n=+1$ order is large.  In the short excitation pulse limit $(4\omega_R\mp\Delta_L)\tau<<1$, thus both $n=\pm 1$ orders contribute equally.  This conclusion is consistent with the notion \cite{schneble1} that the frequency bandwidth of a long pump pulse is too narrow to excite the $n=-1$ order because of the large frequency mismatch (dotted curve, Fig. 1c), whereas the bandwidth of a short pulse is sufficient to overlap both $n=\pm 1$ orders simultaneously \cite{note2}.  

To investigate optical field generation and propagation dynamics, we examine the Maxwell equation obeyed by the generated optical field that counter-propagates the pump field, that is 
($\kappa_{0}=2\pi\omega_{B}|d_{12}|^2/(c\hbar)$)
\begin{eqnarray}
\frac{\partial E_{B}^{(+)}}{\partial x}+\frac{1}{c}\frac{\partial E_{B}^{(+)}}{\partial t}=-i\frac{\kappa_0}{\Delta}E_{B}^{(+)}-i\frac{\kappa_{0}}{\Delta}E_{L}^{(+)}(P_{-1}+P_{+}),
\end{eqnarray}
where $P_{\pm 1}=\psi_{0}\psi_{\pm 1}^{*}\,e^{i(4\omega_{R}\mp\Delta_{L})t}$. As we show below, the inclusion of $n<0$ orders in the source term significantly alters the propagation dynamics of the generated field in the short-pump-pulse limit.

Differentiating $P_{\pm 1}$ with respect to time and assuming a non-depleted ground state
condensate, we obtain

\begin{eqnarray}
\frac{\partial P_{\pm 1}}{\partial t}=D_{\pm}P_{\pm 1}
+ig_{0}\delta\frac{E_{B}^{(\pm)}}{E_{L}^{(\pm)}}|\psi_{0}|^2,
\end{eqnarray}
where $D_{\pm}=i(4\omega_{R}\mp\Delta_{L})-(4R+\gamma_B)$.

\begin{figure}
\centering
\includegraphics[angle=90,width=3in]{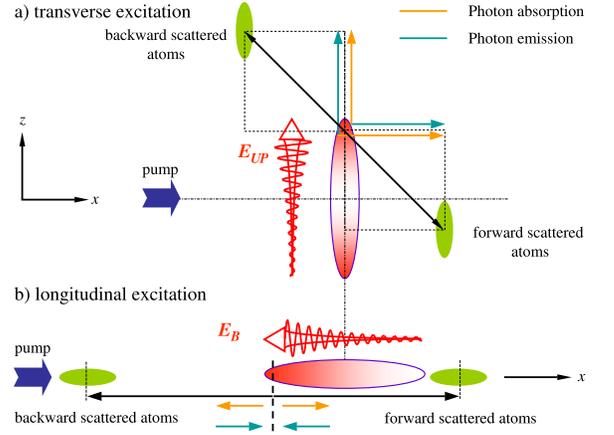} 
\caption{Growth and propagation of electromagnetic wave by stimulated generation of the backward (red arrow) spontaneously
emitted photon. (a) Transverse excitation scheme (only the up-mode is shown) showing the paired forward- and backward-scattered condensates. (b) Longitudinal excitation scheme  (only the backward mode is shown) showing the paired forward- and backward-scattered condensates.}
\end{figure}

Taking the time Fourier transform of Eqs. (2) and (3), and noting that
for strong, short-pulsed excitation $4R\tau>>1$, we obtain in Fourier space ($n_{0}=\sum_{n}|\psi_{n}|^2$),

\begin{eqnarray}
\frac{\partial\epsilon^{(+)}}{\partial x}&=&\left[-i\frac{\kappa_0}{\Delta}n_0+i\frac{\omega}{c}\right]\epsilon^{(+)}\nonumber\\
&+&\kappa_0g_0n_0\left[\frac{1}{-i\omega-D_{+}}-\frac{1}{-i\omega-D_{-}}\right]\epsilon^{(+)}.\quad
\end{eqnarray}
where $\epsilon^{(+)}$ is the Fourier transform of $E_B^{(+)}$, $\omega$ is the transform variable, and the
matter-optical wave phase matching condition $\Delta K=k_M-(k_L+k_B)=0$ has been enforced ($k_M$ is the matter wave vector).
Note that for both long- and short-pulse excitations we have $|D_{\pm}\tau|>>1$, thus
with the assumption of an initial pulse shape of $I_B(x,t)=I_B(0,0)e^{-t^2/\tau^2}$, Eq. (4) yields:
\cite{note1b} 

\begin{eqnarray}
I_{B}(x,t)=I_{B}(0,0)e^{(G-\beta)x}\exp\left[-2\left(\frac{t}{\tau}-\frac{x}{V_g\tau}\right)^2\right],
\end{eqnarray}
where the intensity $I_B\propto |E_B|^2$, and $\beta=2\kappa_{0}\Gamma_{0}n_{0}/|\Delta|^2$ is the linear field loss coefficient. The small-signal propagation gain and group velocity are given by ($\alpha=D_{+}/D_{-}$)

\begin{eqnarray}
G\approx\frac{\kappa_{0}g_{0}n_{0}}{-D_{+}}(1-\alpha),\quad
\frac{1}{V_g}\approx\frac{1}{c}
+\frac{\kappa_{0} g_{0}n_{0}}{D_{+}^2}(1-\alpha^2).
\end{eqnarray}

In the short-pump-pulse limit, we have 
$|(4\omega_{R}\pm\Delta_{L})\tau|<<1$ and both $n=\pm 1$ orders are equally important.  
Since $|(4R+\gamma_B)\tau|>>1$, we have $\alpha\approx 1$.

Three consequences of this are: (1) the stimulated generated electromagnetic field propagates with group velocity $V_g\approx c\,$; (2) the gain is small; and (3) the superradiantly generated field is significant only when it reaches the opposite end of the medium after a full-length propagation.  We thus conclude that under short-pulse excitation the gain is not adequate for efficient atom scattering in the early stage of propagation and most superradiant scattering events occur near the exit 
end of the condensate \cite{note2b}.  This provides a clear explanation for the apparent ``asymmetry" observed in \cite{schneble1}.  Indeed,
with the radiation source located at the end, which must be taken as the reference point, there is no asymmetry \cite{zobay, note2ba} [it is incorrect to take a vertical plane passing the center of the condensate as the reference plane. See
Fig. 2].  Second, under short-pulse excitation there will be no significant group velocity reduction.  This unimpedated propagation velocity of the generated field also has important consequences.  The fast propagation, together with the small size of the condensate, now
impose a stringent requirement of a large one-photon detuning in order to validate the adiabatic elimination of the upper electronic state.  That is, one must have $|\delta|>>\gamma_{photon}\approx c/L>10^{12}$ \cite{note2bb}.  Without such large detunings, as in the case of all short-pulse experiments reported to date, non-adiabatic corrections to the system dynamics are significant.  In fact, the significant spread of the atomic cloud and halos observed in Ref.\cite{schneble1} are direct consequences of this non-adiabaticity which leads to significant excitation of the upper electronic state and results in severe spontaneous emission.  We note that it is precisely this
non-adiabacity-resulted absorptions of the forward pump and the (weak) reflected pump by the exit facet of the cell \cite{note2c} that give atoms one photon
recoil momentum, resulting in a forward (thicker) and a backward (fainter)
atom halo by the subsequent spontaneous emission \cite{kozuma}.  Thus, in the short-pulse limit with the one-photon detunings reported to date, ``matter-wave superradiant scattering" has lost its meaning as there are a large number of atoms spread in various momentum states with random phases. 

Finally, we examine the effect due to the depletion of the ground-state condensate, which we assume to be small
but not negligible.  We show that a sensible understanding of this case can still be gained, and the significant pulse 
front-edge steepening for the generated field can be qualitatively predicted reasonably well.

To this end we start with equations of motion for the field amplitude, polarization $P_{+1}$ and population difference $Z=|\psi_{1}|^2-|\psi_{0}|^2$ as \cite{note3}

\begin{subequations}
\begin{eqnarray}
&&\frac{\partial E_{B}^{(+)}}{\partial x}-\frac{1}{c}\frac{\partial E_{B}^{(+)}}{\partial t}=-i\frac{\kappa_0}{\Delta}n_0 E_{B}^{(+)}-i\frac{\kappa_{0}}{\Delta}E_{L}^{(+)}P_{+1},\quad\quad\\
&&\frac{\partial P_{+1}}{\partial t}=-2\Gamma P_{+1}
+ig_{0}\delta\frac{E_{B}^{(+)}}{E_{L}^{(+)}}Z,\\
&&\frac{\partial Z}{\partial t}=-2\Gamma Z
+i2g_{0}\delta\left[\frac{E_{B}^{(-)}}{E_{L}^{(-)}}P_{+1}-c.c.\right],
\end{eqnarray}
\end{subequations}
where $\Gamma=4R+\gamma_B$.  In general, Eqs. (7a-7c) can only be solved numerically. 
However, insight into the pulse front steepening behavior can be gained by
considering the following weak excitation where $I_{B}/I_{L}<10^{-6}$.)
Thus, the leading contribution in this regime to the population difference is the first term on the 
right hand side of Eq. (7c),
making $Z\approx n_{0}e^{-2\Gamma t}=n_0\sum_{l}^{\infty}(-1)^{l}(2\Gamma t)^{l}/l!$. Equation (7a) now reads,
in the Fourier space,
\begin{eqnarray}
\frac{\partial\epsilon}{\partial x}=-i\left[\beta+\frac{\omega}{c}\right]\epsilon+\frac{\kappa_{0}g_{0}n_{0}}{i\omega+2\Gamma}\left[1+\sum_{l=1}^{\infty}\frac{(-i2\Gamma)^l}{l!}\frac{\partial^l}{\partial\omega^l}\right]\epsilon,
\end{eqnarray}
where the derivative terms on the right side are the contributions due to small ground state depletion.  It is known that the Fourier transform of a product of a time polynomial and a Gaussian function
yields a product of a frequency polynomial of the same degree and a Gaussian function in Fourier space.
To explain the front-edge steepening of the generated field it is sufficient to keep the first few terms.  Thus, Eq. (8), after applying the inverse transform, yields a solution that is a product of a Gaussian function and a time polynomial function.  In our case this type of solution pushes the peak of the Gaussian function forward because of the leading term in the polynomial has a negative coefficient. 

\begin{figure}
\centering
\includegraphics[angle=90,width=3in]{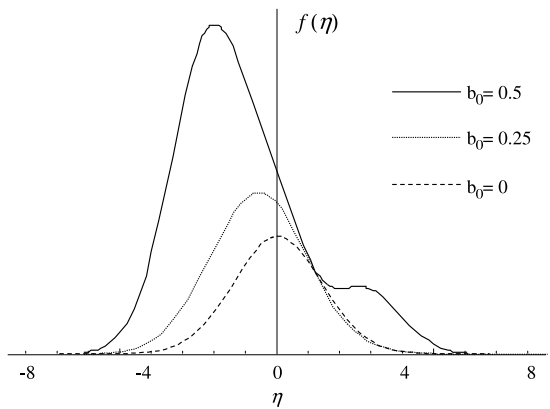} 
\caption{Time domain plot of the product of a Gaussian and a 4th-order time polynomial,
$f(\eta)=(1+b_0)\,e^{-\eta^2/4}\sum_{l=0}^4(-b_0\eta)^l$, with a linear gain dependent coefficient $b_0$ where the coefficient of the first non-constant term is negative (as in our case).  
}
\end{figure} 

In Fig. 3 we have plotted
the absolute value of the field amplitude in the time domain by neglecting $l>4$ terms before performing the inverse transform.  As expected, the peak of the Gaussian pulse profile moves forward, exhibiting steepening of the
front edge of the generated field profile as the excitation rate increases.  This behavior is in 
qualitative agreement 
with experiments and numerical simulations in both longitudinal and transverse excitation geometries
\cite{inouye1, hilliard}, showing our theory captures the essential physics of the 
generated-field-front-edge-steepening effect and peak-forward-shifting effect. We emphasize that regardless of the excitation geometries, the essential propagation characteristics such as ultra-slow propagation with substantial gain (for long pulses) and fast propagation with small gain (for short pulses) will remain.

It should be noted that the present work has not included the atomic mean field contributions.  Recent studies \cite{lunew}
have shown that the inclusion of mean field contribution and boson exchange interaction can lead to very intriguing and important
new effects in the long pulsed pump field.  

In conclusion, we have provided the first uniform treatment of long- and short-pulse, matter-wave superradiance scattering processes.  In the long-pump-pulse limit we have revealed the full propagation dynamics of the generated optical field that has exponential growth and ultra-slow propagation characteristics.  We have also shown that the pulse front-steepening effect of the generated field when the excitation rate increases can be satisfactorily
explained by depletion of the ground state condensate.  In the short-pump-pulse limit, this unified theory predicts unimpeded light propagation and it can explain all experimental observations reported to date including at-end radiation, the apparent ``asymmetry" between the forward- and backward-scattered atoms, the significant atom cloud and the forward-backward halos generated due to significant non-adiabatic processes and fast propagation.  The analytical theory present here can, for the first time, explain the full propagation dynamics of the matter-wave superradiant scattering process in both the long- and short-pulse regimes.


\begin{thebibliography}{13}
\bibitem{note0b} Photons having wave vectors subtend large angles with respect to the long axis experience negligible gain.  Processes involving the forward-generated field (i.e., in the direction of the pump laser) do not lead to appreciable net momentum transfer, and hence are neglected.
\bibitem{note0} Under short-pulse excitation, adequate one-photon detunings should be such that $|\delta|>>\gamma_{photon}\approx 10^{12}$ Hz.
\bibitem{note2} Because of momentum conservation $n<0$
orders appear only after $n>0$ order scatterings have occurred and sufficient number of backward photons have been generated.
\bibitem{note1b} For mathematical simplicity only the first order Taylor expansion of $1/(-i\omega-D_{\pm})$ is kept.
\bibitem{note2b} This is precisely why, unlike the long pulse case, the shapes of the scattered ``condensate" in the short-pulse limit do not resemble that of the original condensate. This can be clearly seen in Refs.\cite{inouye1,schneble1} but has never been
explain by any theories.
\bibitem{note2ba} This end-radiation as the cause of the apparent ``asymmetry" was speculated in Ref.\cite{schneble1}.  Here, we provide a concrete mathematical explanation.
\bibitem{note2bb} We emphasize that the concept of fast photon relaxation usually is not correct unless for short pulse experiments.
\bibitem{note2c} This reflected field causes weak back-scattering of atoms and
it occurs mostly near the center of the condensate.  
Back-scattered atoms by reflected-pump can be seen even
in the long-pulse excitation limit (see Fig. 1B of Ref.\cite{schneble1}).
\bibitem{note3} We neglect higher-order scatterings and $n<0$ orders but allow depletion of the ground state condensate.
High order contributions will be treated in separate studies and published elsewhere.
\end{thebibliography}
\end{document}